\documentstyle[aps]{revtex}
\input epsf

\begin{document}
\title{Electric field induced by magnetic flux motion in superconductor containing
fractal clusters of a normal phase}
\author{Yu. I. Kuzmin}
\address{Ioffe Physical Technical Institute of Russian Academy of Sciences,\\
Polytechnicheskaya 26 St., Saint Petersburg 194021 Russia\\
e-mail: yurk@shuv.ioffe.rssi.ru; iourk@mail.ru\\
tel.: +7 812 2479902; fax: +7 812 2471017}
\date{\today}
\maketitle

\begin{abstract}
The influence of the fractal clusters of a normal phase, which act as
pinning centers, on the dynamics of magnetic flux in percolative type-II
superconductor is considered. The voltage-current characteristics of such a
superconductor are obtained taking into account the effect of fractal
properties of cluster boundaries on the magnetic flux trapping. It is
revealed that the fractality reduces the electric field arising from
magnetic flux motion and thereby raises the critical current of a
superconductor.\bigskip
\end{abstract}

One of the most promising ways of increasing the current-carrying capability
of superconductors is to make the artificial pinning centers in their bulk.
Specifically, the normal phase clusters created in the course of the film
growth at the sites of defects on the boundary with the substrate can act as
such centers \cite{beasley}-\cite{kuzminpss}. New possibilities for pinning
enhancement are opening in the case that the normal phase clusters have
fractal boundaries \cite{kuzminpla}, \cite{kuzmintpl}. It is the influence
of such fractal clusters on the critical currents and on the voltage-current
({\it V-I}) characteristics of the superconductor in the resistive state
that will interest us.

The problem setting is quite similar to that considered in Ref.~\cite
{kuzminpla}. We deal with the superconductor containing inclusions of a
normal phase of columnar shape. In the course of the cooling below the
critical temperature in the magnetic field (``field-cooling'' regime) the
two-dimensional distribution of the trapped magnetic flux is created in such
a superconducting structure. As this takes place, the magnetic flux is
locked in the isolated normal phase clusters that gives rise to effective
pinning. Then the transport current is passed through the sample
transversely to the orientation of the frozen magnetic field. Suppose that
there is a superconducting percolation cluster in the plane carrying the
electric current. When the transport current is rising, the trapped flux
remains unchanged until the vortices start to break away from the clusters
of pinning force weaker than the Lorentz force created by the current. Thus,
the vortices will cross the superconducting space through the weak links,
which connect the normal phase clusters between themselves. Such weak links
form especially readily at the sites of various structural defects in
high-temperature superconductors (HTS), which are characterized by an
extremely short coherence length \cite{mezzetti}, \cite{scalapino}-\cite
{sonier}. In conventional low-temperature superconductors weak links can be
formed due to the proximity effect in sites of minimum distance between the
adjacent normal phase clusters.

Thus, whatever the microscopic nature of weak links may be, they form the
channels for vortex transport. In accordance with their configuration each
normal phase cluster contributes to the overall critical current
distribution. When a transport current is gradually increased, the vortices
will break away first from the clusters of small pinning force, and
therefore, of small critical current. Thus the decrease in the trapped
magnetic flux $\Delta \Phi $ is proportional to the number of all the normal
phase clusters of critical currents less than a preset value $I$. Hence, the
relative decrease in the trapped flux can be expressed with the cumulative
probability function $F=F(I)$ for the distribution of the critical currents
of clusters: 
\begin{equation}
\frac{\Delta \Phi }{\Phi }=F(I)\text{ \ \ \ \ \ , \ \ \ \ \ \ \ \ \ \ \ \ \
\ \ \ \ \ \ \ \ where \ \ \ \ \ \ \ \ \ \ \ \ \ \ \ \ \ \ \ \ \ \ \ \ \ \ }%
F(I)={\bf Pr}\left\{ \forall I_{j}<I\right\}  \label{probi}
\end{equation}
The right-hand side of Eq.~(\ref{probi}) is the probability that any $j$th
cluster has the critical current $I_{j}$ less than a given upper bound $I$.

On the other hand, the magnetic flux trapped into a single cluster is
proportional to its area $A$, so the decrease in the total trapped flux can
be represented by the cumulative probability function $W=W(A)$ for the
distribution of the areas of the normal phase clusters, which is a measure
of the number of the clusters of area smaller than a given value of $A$:

\begin{equation}
\frac{\Delta \Phi }{\Phi }=1-W(A)\text{ \ \ \ \ \ , \ \ \ \ \ \ \ \ \ \ \ \
\ \ \ \ \ \ where \ \ \ \ \ \ \ \ \ \ \ \ \ \ \ \ \ \ \ \ \ \ \ \ }W(A)={\bf %
Pr}\left\{ \forall A_{j}<A\right\}  \label{proba}
\end{equation}

In order to clear up how the transport current acts on the trapped magnetic
flux, it is necessary to find out the relationship between the distribution
of the critical currents of the clusters (Eq.~(\ref{probi})) and the
distribution of their areas (Eq.~(\ref{proba})). This problem was solved in
Ref.~\cite{kuzminpla} for the case of the exponential distribution of the
areas of the normal phase clusters with fractal boundary, which occurs in
HTS structures based on YBCO films. The exponential distribution is the
special case of gamma distribution, which describes the cluster area
distribution in the most general way: 
\begin{equation}
W(A)=%
%TCIMACRO{\dfrac{\gamma (g+1,\frac{A}{A_{0}})}{\Gamma (g+1)} }%
%BeginExpansion
{\displaystyle{\gamma (g+1,\frac{A}{A_{0}}) \over \Gamma (g+1)}}%
%EndExpansion
\label{ardis}
\end{equation}
where $\gamma (\nu ,z)$ is the incomplete gamma function, $\Gamma (\nu )$ is
Euler gamma function, $A_{0}$ and $g$ are the parameters of gamma
distribution which set the mean area of the cluster $\overline{A}=A_{0}(g+1)$
and its standard deviation $\sigma _{A}=A_{0}\sqrt{g+1}$. Gamma distribution
of Eq.~(\ref{ardis}) is reduced to the exponential one at $g=0.$

Using the same expression for the critical current of the fractal cluster $%
I=\alpha A^{-D/2}$ (where $\alpha $ is the form factor, $D$ is the fractal
dimension of the cluster perimeter, or so-called coastline dimension \cite
{mandelbrotfcd}, \cite{mandelbrotgn}) as in Ref.~\cite{kuzminpla}, and
taking into account the initial relationship of Eq.~(\ref{probi}), we can
get the distribution of the critical currents in the general case of
gamma-distributed cluster areas: 
\begin{equation}
F(i)=\frac{\Gamma (g+1,Gi^{-\frac{2}{D}})}{\Gamma (g+1)}  \label{curdis}
\end{equation}

\[
\text{where \ }G\equiv \frac{\theta ^{\frac{2}{D}(g+1)+1}}{\left( \theta
^{g+1}-\frac{D}{2}e^{\theta }\Gamma (g+1,\theta )\right) ^{\frac{2}{D}}}%
\text{ \ \ \ \ \ \ \ \ , \ \ \ \ \ \ \ \ \ \ \ \ \ \ \ \ \ \ \ \ \ \ \ \ \ }%
\theta \equiv \frac{D}{2}+g+1 
\]
$\Gamma (\nu ,z)=\Gamma (\nu )-\gamma (\nu ,z)$ is the complementary
incomplete gamma function, $i\equiv I/I_{c}$ is the dimensionless transport
current, and $I_{c}=\alpha \left( A_{0}G\right) ^{-D/2}$ is the critical
current of the resistive transition.

The found cumulative probability function of Eq.~(\ref{curdis}) provides the
comprehensive description for the effect of the transport current on the
trapped magnetic flux. Using this function, the probability density $%
f(i)\equiv dF/di$ for the critical current distribution can be readily
derived: 
\begin{equation}
f(i)=\frac{2G^{g+1}}{D\Gamma (g+1)}i^{-\frac{2}{D}(g+1)-1}\exp \left( -Gi^{-%
\frac{2}{D}}\right)  \label{densi}
\end{equation}
This function is normalized to unity over all possible positive values of
the critical currents.

The resistive state comes in the range of the currents $i>1$, when the
magnetic flux motion gives rise to the voltage across a superconductor. The
appearance of some finite resistance causes the energy dissipation to
accompany the passage of electric current. As for any type-II superconductor
with pinning centers the dissipation in the resistive state does not mean
the destruction of phase coherence yet. The superconductivity does not fully
collapse until a growth of dissipation becomes avalanche-like as a result of
thermo-magnetic instability.

In the resistive state the superconductor is adequately specified by its 
{\it V-I} characteristic. The critical current distribution of Eq.~(\ref
{densi}) allows us to find the electric field arising from the magnetic flux
motion after the vortices have been broken away from the pinning centers.
Inasmuch as each normal phase cluster contributes to the total critical
current distribution, the voltage across a superconductor $V=V(i)$ is the
response to the sum of effects made by each cluster: 
\begin{equation}
V=R_{f}\int\limits_{0}^{i}(i-i^{\prime })f(i^{\prime })di^{\prime }
\label{volt}
\end{equation}
where $R_{f}$ is the flux flow resistance. The similar approach is used
universally to consider behavior of the clusters of pinned vortex filaments 
\cite{warnes}; to analyze the critical scaling of {\it V-I} characteristics
of superconductors \cite{brown}; that is to say, in all the cases where the
distribution of the depinning currents occurs. The following consideration
is primarily concentrated on the consequences of the fractal nature of the
normal phase clusters specified by the distribution of Eq.~(\ref{densi}), so
all the problems related to possible dependence of the flux flow resistance $%
R_{f}$ on a transport current will not be taken up here.

After the substitution of the probability density function of Eq.~(\ref
{densi}) in Eq.~(\ref{volt}), upon integration, the voltage across a
superconductor can be written in its final form: 
\begin{equation}
\frac{V}{R_{f}}=i+\frac{DG^{g+1}i^{-\frac{2}{D}(g+1)+1}}{2\left( g+1-\frac{D%
}{2}\right) \Gamma \left( g+2\right) }\,_{2}F_{2}\left( \left. 
\begin{array}{cc}
g+1, & g+1-\frac{D}{2} \\ 
g+2, & g+2-\frac{D}{2}
\end{array}
\right| -\frac{G}{i^{\frac{2}{D}}}\right) -\frac{\Gamma \left( g+1-\frac{D}{2%
}\right) }{\Gamma \left( g+1\right) }G^{\frac{D}{2}}  \label{voltgen}
\end{equation}
where $_{2}F_{2}\left( \left. 
\begin{array}{cc}
\nu , & \mu \\ 
\xi , & \eta
\end{array}
\right| z\right) $ is the generalized hypergeometric function.

In the special case of an exponential distribution (at $g=0$) the expression
of Eq.~(\ref{voltgen}) can be simplified: 
\begin{equation}
\frac{V}{R_{f}}=i\exp \left( -Ci^{-\frac{2}{D}}\right) -C^{\frac{D}{2}%
}\Gamma \left( 1-\frac{D}{2},Ci^{-\frac{2}{D}}\right) \text{ \ \ \ , \ \ \ \
\ where \ \ \ \ \ \ \ \ \ }C\equiv \left( \frac{2+D}{2}\right) ^{\frac{2}{D}%
+1}  \label{voltspec}
\end{equation}

In the extreme cases of Euclidean clusters ($D=1$) and clusters of boundary
with the maximum fractality ($D=2$) the formula of Eq.~(\ref{voltspec}) for
the voltage across a sample can be further transformed: 
\begin{equation}
\left( D=1,\,g=0\right) \text{: \ \ \ \ \ \ \ \ \ \ \ \ \ \ \ \ \ \ \ }\frac{%
V}{R_{f}}=i\exp \left( -\frac{3.375}{i^{2}}\right) -\sqrt{3.375\pi }\text{%
erfc}\left( \frac{\sqrt{3.375}}{i}\right)  \label{volteuclid}
\end{equation}
\begin{equation}
\left( D=2,\,g=0\right) \text{: \ \ \ \ \ \ \ \ \ \ \ \ \ \ \ \ \ \ \ \ \ \
\ \ \ \ \ \ \ \ \ \ \ \ \ \ \ \ \ \ \ }\frac{V}{R_{f}}=i\exp \left( -\frac{4%
}{i}\right) +4%
%TCIMACRO{\func{Ei}}%
%BeginExpansion
\mathop{\rm Ei}%
%EndExpansion
\left( -\frac{4}{i}\right)  \label{voltpeano}
\end{equation}
where erfc$(z)$ is the complementary error function, $%
%TCIMACRO{\func{Ei}}%
%BeginExpansion
\mathop{\rm Ei}%
%EndExpansion
\left( z\right) $ is the exponential integral function.

\begin{figure}[tbp]
\epsfbox{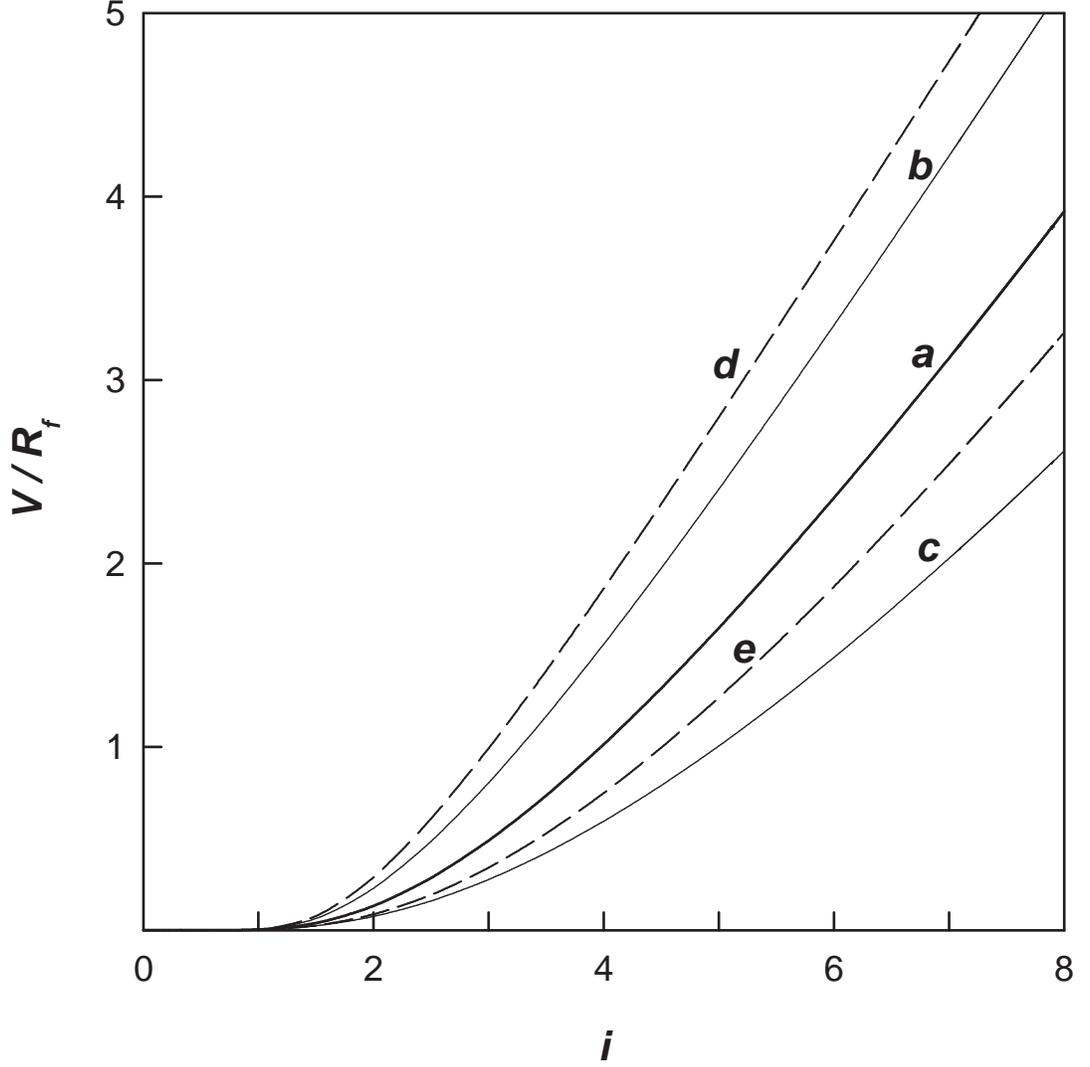}
\caption{Voltage-current characteristics of superconductor containing
fractal clusters of a normal phase. The solid lines (a), (b), and (c)
pertain to the exponential distribution of the cluster areas ($g=0$). Curve
(a) corresponds to the case of fractal clusters of dimension $D=1.44$; curve
(b) -- to Euclidean clusters ($D=1$); curve (c{}) -- to clusters of boundary
with the maximum fractality ($D=2$). Two dotted lines (d) and (e) are given
for the case of gamma distribution at $g=0.5$. Curve (d) corresponds to
Euclidean clusters ($D=1$), curve (e) -- to clusters of the most fractality (%
$D=2$).}
\label{figure1}
\end{figure}

\begin{figure}[tbp]
\epsfbox{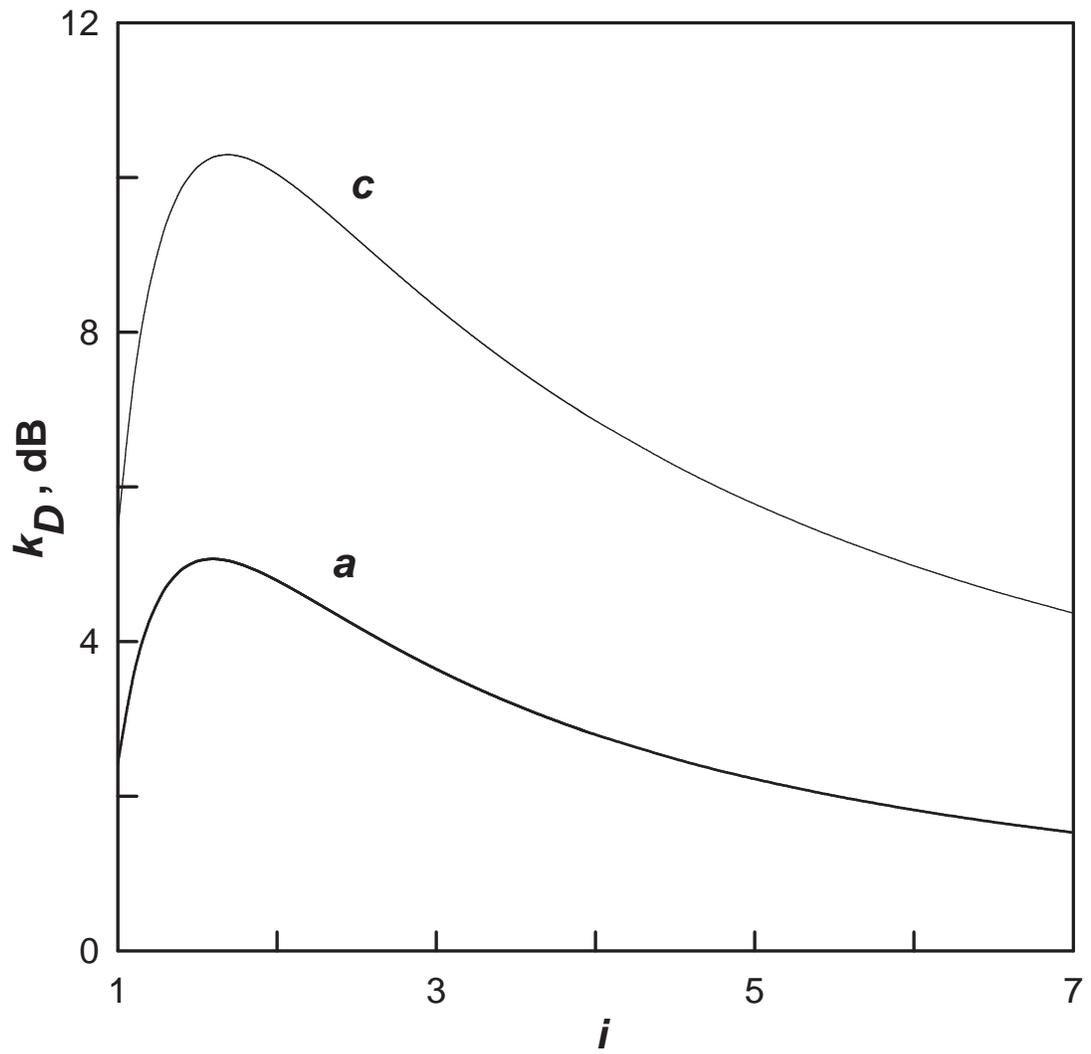}
\caption{The pinning gain for the cases: (a) -- $D=1.44,\,g=0$, and (c) -- $%
D=2,\,g=0$.}
\label{figure2}
\end{figure}

\begin{figure}[tbp]
\epsfbox{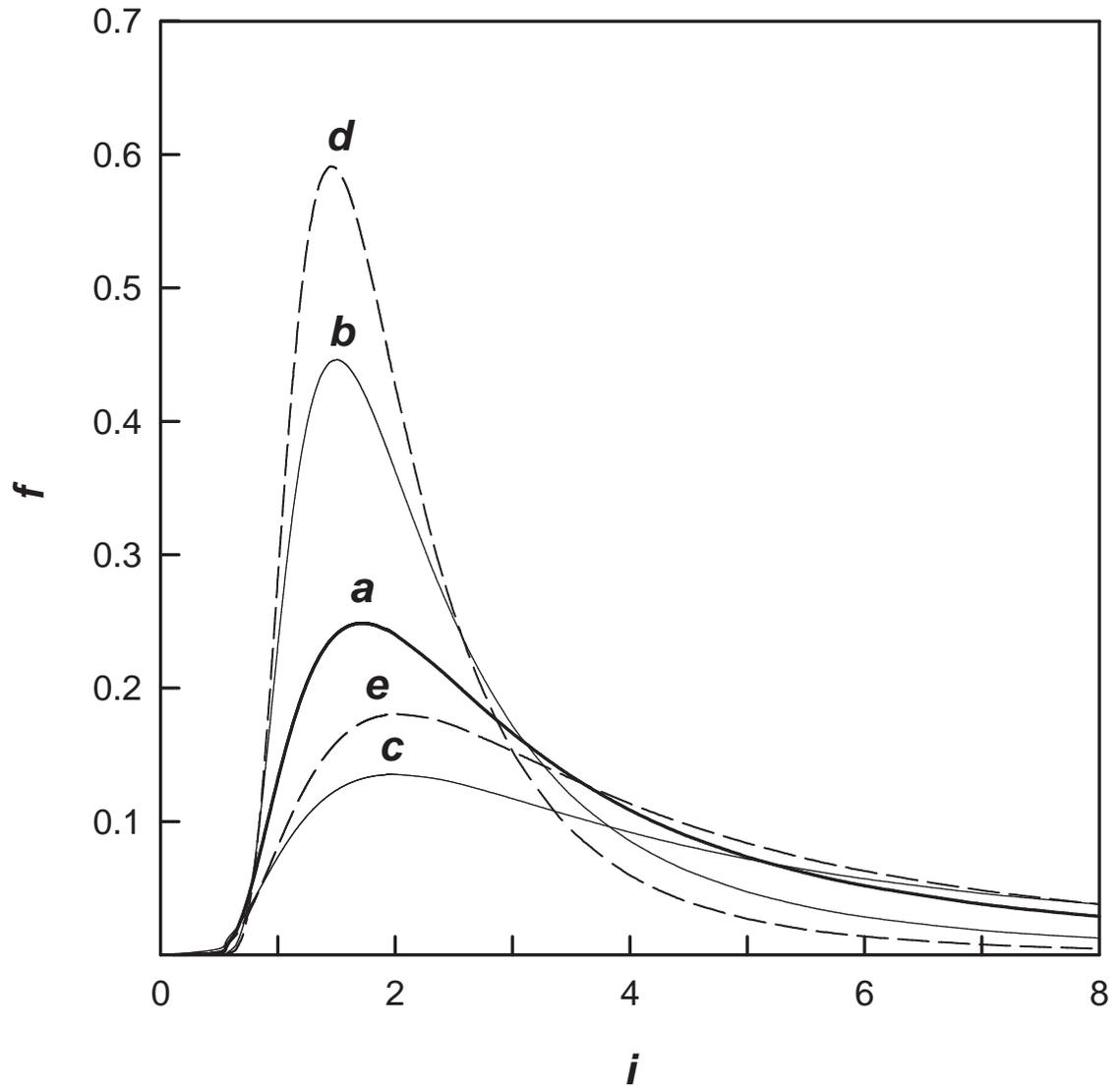}
\caption{Influence of the fractal dimension of the cluster boundary on the
critical current distribution. Curve (a) corresponds to the case of $%
D=1.44,\,g=0$; curve (b) -- to $D=1,\,g=0$; curve (c) -- to $D=2,\,g=0$;
curve (d) -- to $D=1,\,g=0.5$; curve (e) -- to $D=2,\,g=0.5$.}
\label{figure3}
\end{figure}

\begin{figure}[tbp]
\epsfbox{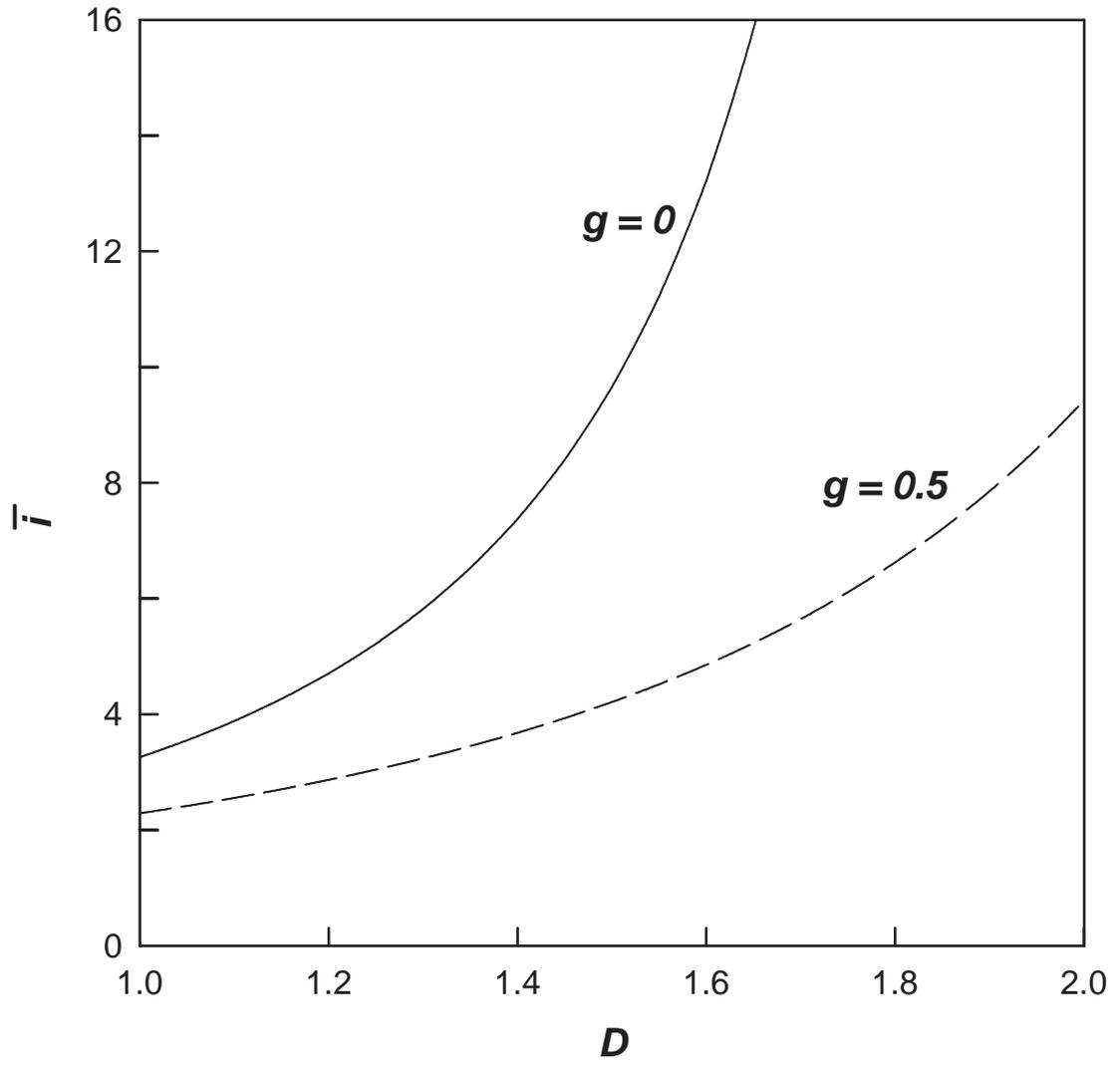}
\caption{The dependence of the average critical current $\overline{i}$ on
the fractal dimension at the different values of gamma distribution
parameter $g$.}
\label{figure4}
\end{figure}

The {\it V-I} characteristics of a superconductor containing fractal
clusters of a normal phase are presented in Fig.~\ref{figure1}. All the
curves are virtually starting with the transport current value of $i=1$ that
is agreed with the onset of the resistive state found with the use of
cumulative probability function \cite{kuzminpla}. At smaller current the
total trapped flux remains unchanged for lack of pinning centers of such
small critical currents, so the breaking of the vortices away has not
started yet. The {\it V-I} characteristic (a) is drawn in Fig.~\ref{figure1}
for the value of the fractal dimension $D=1.44$, which was found earlier in
Ref.~\cite{kuzminpla} for superconducting YBCO film structures with
exponentially distributed cluster areas ($g=0$). Two thin lines (b) and (c),
calculated for extreme cases of Euclidean clusters and clusters of the most
fractality, bound the region the {\it V-I} characteristics can fall within
for any possible values of fractal dimension at $g=0$. Dotted lines (d) and
(e) show the same extreme cases, but at the different value of gamma
distribution parameter: $g=0.5$. This figure demonstrates that the
fractality reduces appreciably an electric field arising from the magnetic
flux motion. Furthermore, this effect shows up most clearly for exponential
area distribution. Figure~\ref{figure2} displays the behavior of pinning
gain factor 
\[
k_{D}\equiv 20\lg \frac{\Delta \Phi \left( D=1\right) }{\Delta \Phi \left(
current\,\,value\,\,of\,\,D\right) },\text{ \ \ \ \ \ \ \ \ \ \ \ \ \ \ \ \
\ \ \ \ \ \ \ \ dB} 
\]
which is equal to relative decrease in the fraction of magnetic flux broken
away from fractal clusters of the boundary dimension $D$ compared to the
case of Euclidean ones \cite{kuzmintpl}. The decrease in the electric field
with increasing fractal dimension is especially strong in the range of
currents $1<i<3$, where the pinning gain also has a maximum. Both these
effects have the same nature, inasmuch as their reason consists in the
peculiarities of fractal distribution of critical currents of Eq.~(\ref
{densi}). Figure~\ref{figure3} shows how the fractal dimension of the
cluster boundary affects the critical current distribution. It is clearly
seen that this bell-shaped curve broadens out, shifting towards greater
magnitudes of current as the fractal dimension increases. This
re-distribution can be described by superlinear dependence of average
critical current $\overline{i}$\ on the fractal dimension, specified by
Euler gamma function (see Fig.~\ref{figure4}): 
\[
\overline{i}=\frac{\Gamma \left( g+1-\frac{D}{2}\right) }{\Gamma \left(
g+1\right) }G^{\frac{D}{2}} 
\]

The highest current-carrying capability of superconductor is achieved for
exponential area distribution (at $g=0$). In this case the smaller clusters,
which are of greater critical current, make the most contribution to the
overall distribution. Thus the mean critical current rises more steeply at $%
g=0$ than it does at $g=0.5$ with increasing the fractal dimension. The mean
value of the critical current for Euclidean clusters at $g=0$ is equal to $%
\overline{i}(D=1)=\left( 3/2\right) ^{3/2}\sqrt{\pi }=3.2562$, whereas for
clusters of the most fractality ($D=2$) this value diverges. As may be seen
from Fig.~\ref{figure3}, an increase of the fractal dimension causes a
significant broadening of the tail of the distribution $f=f(i)$, whereas the
whole area under a curve remains constant. It means that more and more of
small clusters, which can best trap the magnetic flux, are being involved in
the game. Hence the density of vortices broken away from pinning centers by
the Lorentz force is reducing, so the smaller part of a magnetic flux can
flow, creating the smaller electric field. In turn, the smaller the electric
field, the smaller is the energy dissipated when the current passes through
the sample. Therefore, the decrease in heat-evolution, which could cause
transition of a superconductor into a normal state, means that the
current-carrying capability of the superconductor containing such fractal
clusters is enhanced.

Thus, the fractal properties of the normal phase clusters exert an
appreciable effect on the dynamics of the trapped magnetic flux. The crucial
change of the critical current distribution caused by increasing of the
fractal dimension of the cluster boundary forms the basis of this
phenomenon. The most important result is that the fractality of the clusters
strengthens the flux pinning and thereby hinders the destruction of
superconductivity by the transport current. This phenomenon provides the
principally new possibility for increasing the critical current value of
composite superconductors by optimizing their geometric morphological
properties.

\newpage

\newpage

\newpage

\newpage

\end{document}